\documentclass[twocolumn, prb,superscriptaddress]{revtex4}       
\usepackage{color,graphicx}                               
\usepackage{amssymb}                                      
\usepackage{lineno}                                            

\begin{document}
\title{Non-collinear spin-density wave antiferromagnetism in FeAs}

\author{E. E. Rodriguez}
\affiliation{NIST Center for Neutron Research, NIST, 100 Bureau Dr., Gaithersburg, MD 20878}
\author{C. Stock}
\affiliation{NIST Center for Neutron Research, NIST, 100 Bureau Dr., Gaithersburg, MD 20878}
\affiliation{Indiana University, 2401 Milo B. Sampson Lane, Bloomington, IN, 47408}
\author{K. L. Krycka}
\affiliation{NIST Center for Neutron Research, NIST, 100 Bureau Dr., Gaithersburg, MD 20878}
\author{C. F. Majkrzak}
\affiliation{NIST Center for Neutron Research, NIST, 100 Bureau Dr., Gaithersburg, MD 20878}
\author{P. Zajdel}
\affiliation{Division of Physics of Crystals, Institute of Physics, University of Silesia, Katowice, 40-007, Poland}
\author{K. Kirshenbaum}
\affiliation{Center for Nanophysics and Advanced Materials, Department of Physics, University of Maryland, College Park,
MD 20742}
\author{N. P. Butch}
\affiliation{Center for Nanophysics and Advanced Materials, Department of Physics, University of Maryland, College Park,
MD 20742}
\author{S. R. Saha}
\affiliation{Center for Nanophysics and Advanced Materials, Department of Physics, University of Maryland, College Park,
MD 20742}
\author{J. Paglione}
\affiliation{Center for Nanophysics and Advanced Materials, Department of Physics, University of Maryland, College Park,
MD 20742}
\author{M. A. Green}
\affiliation{NIST Center for Neutron Research, NIST, 100 Bureau Dr., Gaithersburg, MD 20878}
\affiliation{Department of Materials Science and Engineering, University of Maryland, College Park,
MD 20742}

%

\begin{abstract}The nature of the magnetism in the simplest iron arsenide is of fundamental importance in understanding the interplay between localized and itinerant magnetism and superconductivity.  We present the magnetic structure of the itinerant monoarsenide, FeAs, with the B31 structure. Powder neutron diffraction confirms incommensurate modulated magnetism with wavevector $\mathbf{q} = (0.395\pm0.001)\mathbf{c}^*$ at 4 K, but cannot distinguish between a simple spiral and a collinear spin-density wave structure. Polarized single crystal diffraction confirms that the structure is best described as a non-collinear spin-density wave arising from a combination of itinerant and localized behavior with spin amplitude along the $b$-axis direction being (15 $\pm$ 5)\% larger than in the $a$-direction. Furthermore, the propagation vector is temperature dependent, and the magnetization near the critical point indicates a two-dimensional Heisenberg system.  The magnetic structures of closely related systems are discussed and compared to that of FeAs.
\end{abstract}

\maketitle

%


\section{Introduction}

The exact treatment of localized electrons in materials with Fermi surfaces remains a substantial challenge in condensed matter science and has recently come to prominence with the close association of itinerant magnetism and superconductivity in iron based superconductors. A number of iron arsenides, such as LaFeAsO, BaFe$_2$As$_2$, and NaFeAs, exhibit electrical conductivity and antiferromagnetic ordering,\cite{paglione_2010, johnston_2010} but lose such ordering to superconductivity when chemically doped or subjected to high pressure.\cite{kamihara_2008, delaCruz_2008, rotter_2008}  Currently, a central debate on these materials is the origin of the magnetic order, arising from either competing exchange interactions between iron sites or Fermi surface nesting.  The simplest of all iron arsenide systems, the monoarsenide FeAs, may well hold answers to fundamental questions concerning the nature of the Fe--As bond and Fe--Fe interactions in arsenides. The itinerant magnetism of FeAs, with an iron moment $\approx 0.5 \mu_B$, is astonishingly similar to the ground state of compositions associated with iron based superconductors, despite the former having a three-dimensional structural network.  This close relationship warrants further investigation to gain a better understanding of normal state iron arsenides in general, and establish whether superconductivity can be supported in other structural families.

Fundamental to the coexistence of magnetic order and metallic conductivity in this compound is the presence of incommensurate magnetic ordering, which can be interpreted in terms of either a spin density wave (SDW) or a spiral magnetic structure.  The importance of spiral phases in itinerant magnets has been highlighted by recent work on MnSi which unveiled a novel $A$-phase where a unique skyrmion lattice, similar to the vortex phases in superconductors, has been proposed to exist.\cite{muhlbauer_2009, neubauer_2009}  Crucial to this phase is the presence of both a spiral magnetic phase and itinerant electrons, providing yet another motivation to studying itinerant magnets such as FeAs.

\begin{figure}[!b] \centering
\includegraphics[width=0.9\linewidth,angle=0.0]{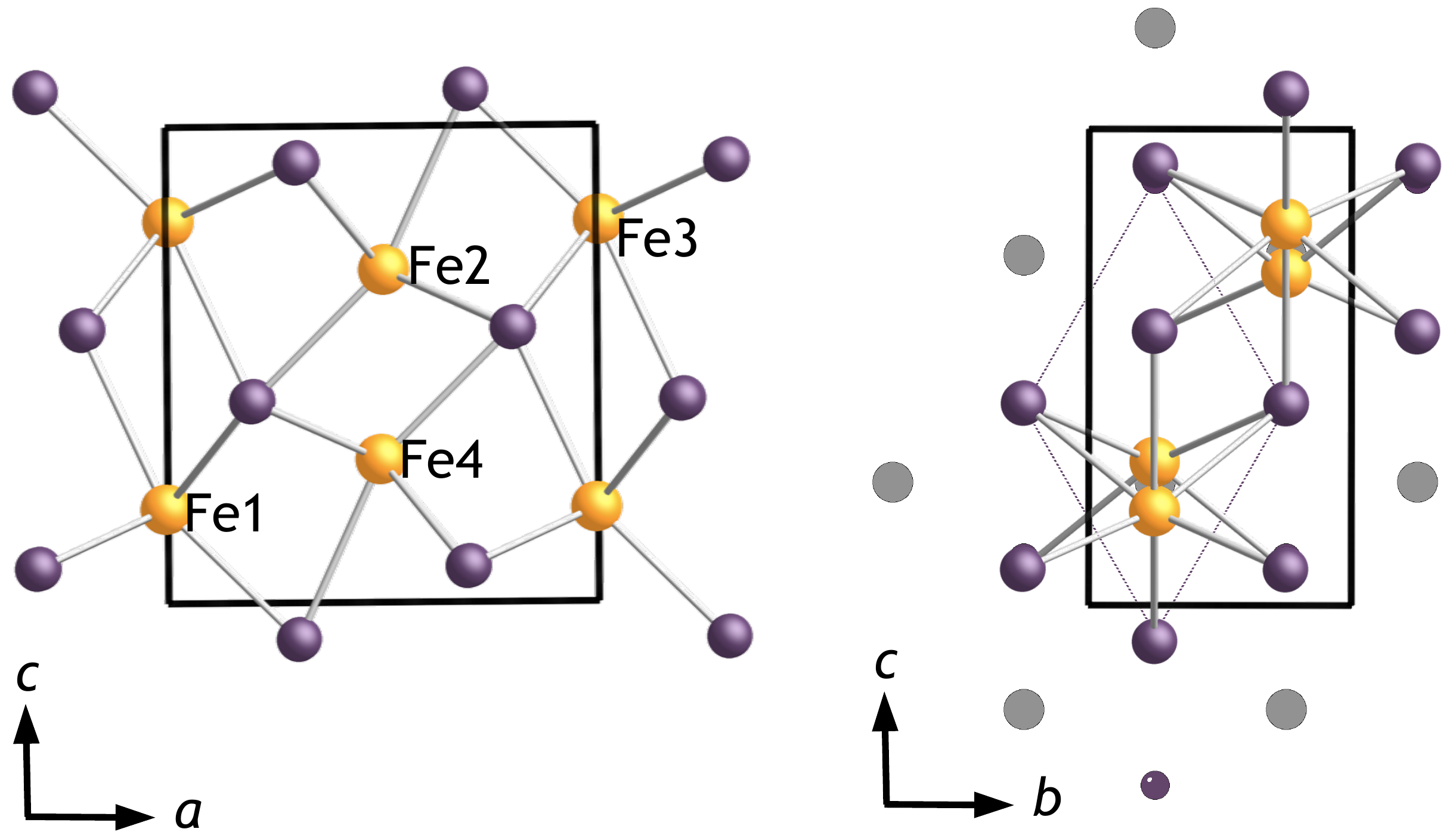}
\caption{[color online] Two views of the orthorhombic crystal structure of FeAs with the unit cell drawn.  For the view down the [010] direction, the four Fe atom sites are labeled.  For the view down the [100] direction, the structure of FeAs is superimposed on the NiAs-type structure (grey atoms) and its hexagonal cell (thin dashed line).  The Fe cations (in orange) are octahedrally coordinated to the arsenic anions (in purple).}
\label{FeAs_structure}
\end{figure}

\begin{figure*}
\includegraphics[width=0.9\linewidth]{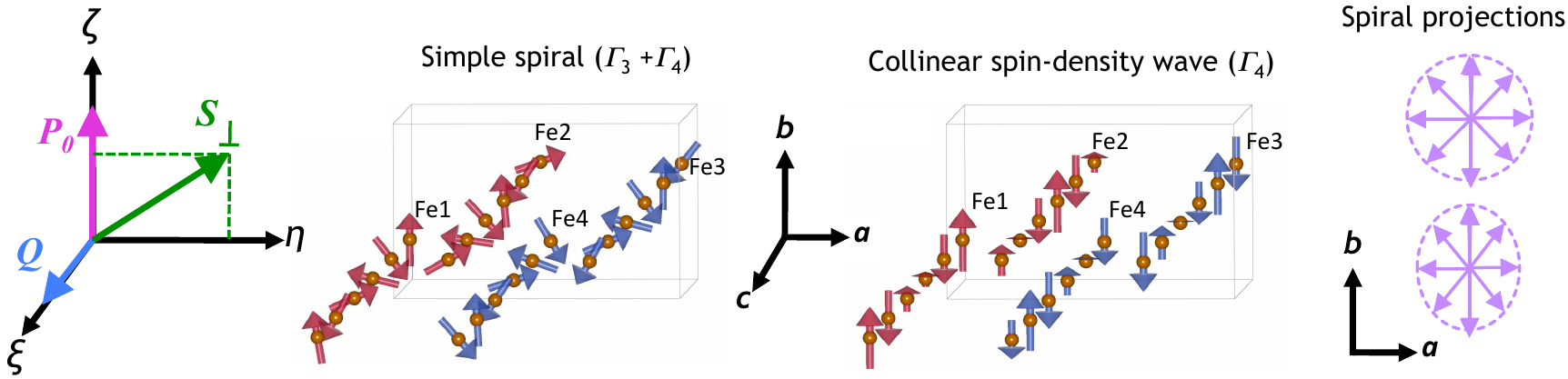}
\caption{[color online] The geometry of the spin polarized neutron experiment with the polarization direction $\mathbf{P}_0$ perpendicular to the scattering vector $\mathbf{Q}$.  Since the magnetic structure factor $\mathbf{F}_M \propto \mathbf{S}_\perp$, in this geometry the spin contribution to the $\zeta$-and $\eta$-axes can be measured.  Illustration of the different modulated magnetic structures in FeAs arising from either a simple spiral or a collinear spin density wave (As atoms left out for clarity).  The simple spiral can be modeled by a combination of representations as $\Gamma_3$ + $\Gamma_4$, whereas the SDW can be modeled by a single representation such as $\Gamma_4$.  Atoms Fe1 and Fe2 belong to one orbit (red arrows), while Fe3 and Fe4 to the other (blue arrows).  The spin projections onto the $ab$-plane for a simple spiral and noncollinear SDW.  The former traces out a circle, and the other an ellipse, with the long-axis along $b$ for FeAs.}
\label{FeAs_magnetism}
\end{figure*}

The B31 structure of FeAs, commonly referred to as the MnP-type structure, can be thought of as a distorted form of the hexagonal NiAs-type structure (See Figure \ref{FeAs_structure}).\cite{rundquist_1962, lyman_1984}  Found in over 400 compounds, the NiAs-type structure occurs frequently for intermetallics combining a transition metal and a metalloid such as Si, As, Se, or Te.\cite{lidin_1998}  FeAs has similar Fe--Fe interactions as the layered FeAs-based superconductors, but is distinguished from the latter by being surrounded by 6 (octahedral) rather than 4 (tetrahedral) arsenic anions.  Metal-metal bonding interactions often lead to crystallographic phase transitions from the hexagonal NiAs-type structure to the orthorhombic MnP-type,\cite{selte_1973, franzen_1974, tremel_1986} as is the case in FeAs (See Figure \ref{FeAs_structure}).

The incommensurate simple spiral model originally proposed for FeAs by Selte \textit{et al.},\cite{selte_1972} has recently come under scrutiny due to the highly anisotropic transport and magnetization properties; for example, susceptibility along the $b$-axis was substantially lower than found along the $a$-axis.\cite{segawa_2009} In addition, Hall coefficient measurements found reentrant sign change with temperature that was assigned to multiple band competition.\cite{segawa_2009}  In this article, we evaluate the magnetic order in FeAs by a series of neutron diffraction experiments and discuss its relation to other itinerant magnetic metals including structurally related transition metal pnictides and various gadolinium compounds.

\section{Experimental}

To investigate the crystal structure and magnetic ordering in FeAs we performed a series of neutron diffraction experiments with both powder and single crystal samples.  The FeAs crystal was grown using the chemical vapor transport technique with a starting powder of FeAs and iodine as the transport agent.  For the powder measurements, we used the BT-1 diffractometer at the NIST Center for Neutron Research (NCNR) with wavelength $\lambda = 2.0782$ \AA (Ge 311 monochromator).  Unpolarized neutron diffraction was performed on the single crystal sample to measure the peak intensity and position of selected magnetic reflections as a function of temperature. These experiments were performed on the BT-9 triple-axis spectrometer with wavelength $\lambda$ = 2.0875 \AA (pyrolytic graphite monochromator).  

Polarized neutron diffraction was performed on the same crystal used in the BT-9 experiments.  The measurements were performed on the SPINS instrument in a configuration where the cold neutron beam with $\lambda = 4.0449$ \AA~was polarized vertically using supermirrors and the crystal aligned so the scattering vector $\mathbf{Q}$ was set perpendicular to the beam polarization direction  $\mathbf{P}_0$. The thin Fe/Si magnetic films within the supermirror reflect spin $+\frac1 2$ neutrons, so only spin $-\frac1 2$ neutrons are transmitted, the latter of which were incident on the sample.  Polarization analysis of the reflected beam was performed with a similar Soller collimator and supermirror assembly in earlier work.\cite{majkrzak_1995}.  Tight collimation following the supermirrors was used to absorb the $+\frac1 2$ neutrons, and flipper coils were then placed before and after the sample, with ($+$) representing the flipper coil on and ($-$) the coil off. 

\section{Results}

\subsection {Unpolarized neutron diffraction}

As determined by neutron powder diffraction data using the FullProf Rietveld program,\cite{fp} FeAs crystallizes in orthorhombic $Pnma$ symmetry with lattice constants $a = 5.45601$(5) \AA, $ b = 3.32843$(3) \AA, and $c = 6.03099$(5) \AA~at 4 K.  The powder measurements also confirm incommensurate modulated magnetism with the wavevector $\mathbf{q} = (0.395\pm0.001)\mathbf{c}^*$, similar to that proposed by Selte \textit{et al}.\cite{selte_1972} The use of colored space groups or Shubnikov groups is insufficient to solve such structures,\cite{bertaut_1962, bertaut_1968} so representational analysis using the program MODY was employed instead.\cite{MODY} The only symmetry elements under which $\mathbf{q}$ remains unchanged are $E$, $C_{2z}$, $\sigma_x$ and $\sigma_y$, which form a little group $G_q$.  The symmetry elements of $G_q$ are presented in matrix form in supplementary information.  With the exception of the identity operator, application of these symmetry elements to the coordinates of the four Fe atoms shows that Fe1 transforms into Fe2 and vice versa by way of a return vector. The same transformations apply for Fe3 and Fe4. Thus, we classify atoms Fe1 and Fe2 as belonging to orbit 1 and Fe3 and Fe4 to orbit 2, as shown in Figure \ref{FeAs_magnetism}. These two orbits constitute the magnetic structure, which  is referred to as a double helical or double spiral structure in past studies of similar systems.\cite{bertaut_1969, felcher_1971, kallel_1974, selte_1979}  

\begin{table}
\caption{The basis functions $\mathbf{\psi}$ for each Fe atom in the unit cell under the four irreducible representations.  The return vector $\epsilon$ is $exp(-i\delta \pi)$ and $\epsilon$* is its complex conjugate.  Here $\delta$ is the $c$-component of the wavevector $\bf{q}$ and is approximately $0.395\pm0.001$ at 4 K.  The fractional coordinates ($x$, $y$, $z$) of the four Fe atoms are shown below the table.}
\begin{tabular}{clcl}
\hline
\hline
Irrep  &    $ \psi$ for orbit 1   &  & $\psi$ for orbit 2  \\
\hline
$\Gamma_1$  & Fe1: (0  1  0)                                           &   & Fe3: (0  1  0)    \\
                           & Fe2: (0 $-\epsilon$ 0)                           &   & Fe4:  (0  $-\epsilon$*  0)\\
\hline
$\Gamma_2$  &  Fe1: (1  0  0); (0 0 1)                                &   & Fe3: (1 0 0); (0 0 1)\\
                           & Fe2: ($-\epsilon$ 0 0); (0 0 $\epsilon$)&   & Fe4: ($-\epsilon$* 0 0); (0 0 $\epsilon$) \\
\hline
$\Gamma_3$  & Fe1: (1  0  0); (0 0 1)                                 &   & Fe3: (1 0 0); (0 0 1)\\
                           & Fe2: ($\epsilon$ 0 0);(0 0 $-\epsilon$) &   & Fe4: ($\epsilon$ 0 0); (0 0 $-\epsilon$) \\
\hline
$\Gamma_4$ & Fe1: (0 1 0)                                                  &   & Fe3: (0 1 0)   \\
                          & Fe2: (0 $\epsilon$ 0)                                 &   & Fe4: (0 $\epsilon$ 0) \\
\hline
\hline
{} &    Fe1: 0.004  0.25   0.199 &{} & Fe3: 0.996 0.75 0.801 \\
{} &    Fe2: 0.496  0.75   0.699 &{} & Fe4: 0.504 0.25 0.301 \\
\end{tabular}
\end{table}

The four symmetry elements in $G_q$ give four irreducible representations, which are all one-dimensional. The representations and their corresponding basis vectors are summarized in Table 1.  The four irreducible representations were then used to fit the magnetic peaks.  A SDW with spin polarization along $b$, $\Gamma_4$, gave a satisfactory fit to the observed powder neutron diffraction data at 4 K, but none of the representations corresponds to a spiral structure.  The combination $\Gamma_4 + \Gamma_3$, however, can reproduce a spiral and gave a similar fit to using $\Gamma_4$ alone. This combination was previously used to model the simple spiral in isostructural MnP,\cite{bertaut_1969} The experimentally determined Fe moments at 4 K are (0.50 $\pm$ 0.05) $\mu_B$ for the simple spiral, whereas the maximum amplitude of the spin polarization in the SDW is (0.58 $\pm$ 0.06) $\mu_B$. The observed and calculated patterns are shown in supplementary information. The powder averaging, however, makes it impossible to distinguish between the two proposed modulated magnetic structures, with identical residuals in both.  

\begin{figure}[!b]
\includegraphics[width=0.9\linewidth,angle=0.0]{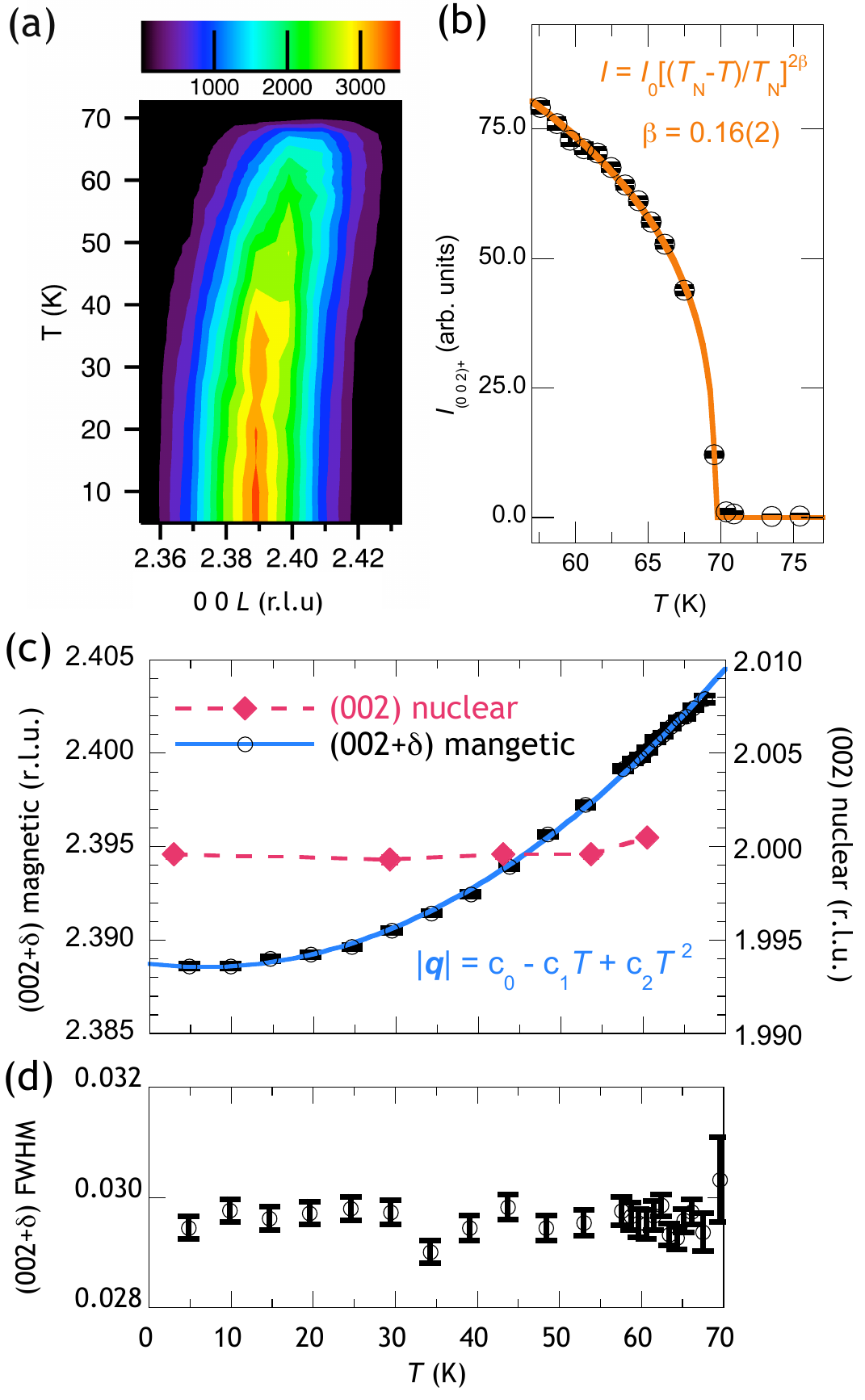}
\caption{[color online] (a) Contour plot of the (0\,\,0\,\,$2+\delta$) magnetic reflection versus temperature. (b) Integrated intensity of the magnetic reflection near the N\`eel point and an order parameter fit to the data.  The critical exponent $\beta$ was found to be $0.16 \pm 0.02$, which is close to a two-dimensional Heisenberg system such as K$_2$MnO$_4$ ($\beta = 0.15 \pm 0.01$)\cite{birgeneau_1973}.  (c) Plot of the center of the magnetic reflection, and therefore the value of the propagation vector $\mathbf{q}$ versus temperature.  A quadratic function fit to the data with the coefficients $c_0 = 0.389$ , $c_1 = -5.360\times10^{-5}$, and $c_2 = 3.991\times 10^{-6}$.  For comparison, the (002) nuclear peak's center is plotted to show the thermal expansion of the lattice, which makes a minimal contribution to the increase of $\mathbf{q}$ with temperature.  (d) Plot of the full width at half maximum (FWHM) of the (0\,\,0\,\,$2+\delta$) magnetic reflection up to the N\`eel point.  All error bars represent an uncertainty of $\pm \sigma$.}\label{FeAs_order}
\end{figure}

From the neutron powder patterns, no crystallographic phase transition was observed in FeAs, raising the possibility that the magnetic phase transition is of second-order.  To characterize the nature of the phase transition, unpolarized neutron diffraction was performed on the single crystal sample to measure the peak intensity and position of selected magnetic reflections as a function of temperature.  The temperature evolution of the (0\,\,0\,\,2$+\delta)$ magnetic reflection is shown in Figure \ref{FeAs_order}.  Near the N\`eel temperature of $T_N = 69.6$(1) K, the critical exponent obtained from a power law fit was found to be $\beta = 0.16 \pm 0.02$ (Figure \ref{FeAs_order}b), which is inconsistent with a three-dimensional Heisenberg ($\approx 0.367$), three-dimensional Ising ($\approx 0.325$), or two-dimensional Ising ($\approx 0.12$) model.\cite{collins}  The critical exponent of FeAs is closer to those of K$_2$NiF$_4$ and K$_2$MnF$_4$, which are $0.1388 \pm 0.004$ and $0.15 \pm 0.01$, respectively.  These systems are best described as two-dimensional Heisenberg models.\cite{birgeneau_1972, birgeneau_1973} 

The present single crystal studies are in disagreement with the temperature dependence of the earlier powder work of Selte \textit{et al}.  The propagation vector, $\mathbf{q}$, derived from the position of the (0\,\,0\,\,2$+\delta$) magnetic reflection increases quadratically up to the N\'eel point (Figure \ref{FeAs_order}c).  Selte \textit{et al.} had reported the propagation vector to have been invariant with temperature and a $T_N$ of 77 K.\cite{selte_1972}  We find $T_N$ = $69.6$(1) K which is much closer to the value obtained by the magnetic susceptibility measurements of Segawa and Ando.\cite{segawa_2009}  Also, the full-width at half maximum (FWHM) of the magnetic reflection does not change as a function of temperature (Figure \ref{FeAs_order}d).

\subsection {Polarized neutron diffraction}

The incoming neutron beam was polarized vertically so that the neutron polarization vector $\mathbf{P_0}$ was set parallel to the $\zeta$-axis and normal to the scattering vector $\mathbf{Q}$.  $\mathbf{S_{\perp}}$ is defined as the component of the spin axis vector perpendicular to $\mathbf{Q}$, and in this experimental configuration, the $\xi$-component of the magnetization is always zero (See Figure \ref{FeAs_magnetism}).  Measuring with two flipper coils, before and after the sample, leads to four possible cross-sections: ($- -$)($+ +$)($-+$) and ($+-$).  The polarized experiment is reduced to measuring the two cross-sections

\begin{eqnarray}
\left( \frac {d\sigma}{d\Omega} \right)_{NSF} & \propto & (b \pm pS_{\perp \zeta})^2 \\
\left( \frac {d\sigma}{d\Omega} \right)_{SF}  & \propto & (\pm pS_{\perp \eta})^2     \\
\end{eqnarray}

\noindent where $b$ is the nuclear scattering length, $p$ is a constant times the magnetic form factor, NSF stands for non-spin flip,  and SF for spin flip.  Therefore, for the nuclear peaks only the NSF cross section is observed since $p$ is zero, and for the magnetic peaks either $S_{\perp \zeta}$ or $S_{\perp \eta}$ are measured since $b$ is zero.  In this polarization geometry, the spin amplitude components along the $\zeta$- and $\eta$-axes are measured, making it straightforward to distinguish between a spiral and a SDW.  A simple spiral with propagation vector along the $c$-direction and moments in the ($ab$)-plane will have equal NSF and SF scattering at a purely magnetic peak ($e.g.$ a magnetic satellite for an incommensurate period).Ê An incommensurate but collinear SDW, also propagating along $c$-direction with moments in the ($ab$)-plane, will not have NSF = SF, except for the special case in which the moments bisect the $a$- and $b$-axes ($i.e.$ at a 45$^{\circ}$ angle).  For the special cases where the SDW is aligned along either $b$ or $a$, intensity in only one channel will be observed.

\begin{figure}[!b] \centering
\includegraphics[width=0.95\linewidth,angle=0.0]{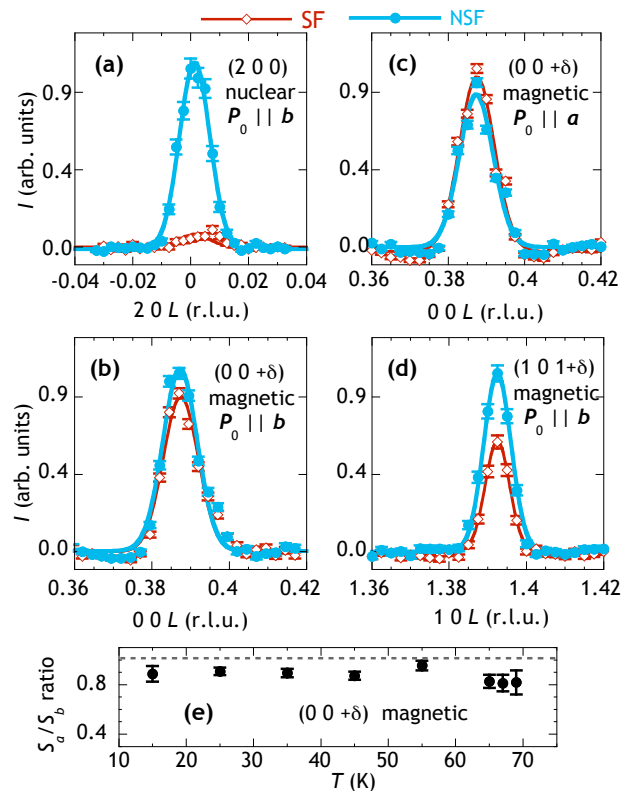}
\caption{[color online] Non-spin flip (NSF) and spin flip (SF) intensities for nuclear and magnetic reflections in FeAs.  In (a) the (2\,\,0\,\,0) nuclear reflection was measured to obtain the flipping ratio, which was found to be $\approx$ 15.  In (b) where $\mathbf{P}_0 \parallel b$-axis, the NSF and SF intensities for the (0\,\,0\,\,$+\delta$) magnetic peak indicate that $S_{\perp \zeta}$ is larger than $S_{\perp \eta}$.  In (c) where $\mathbf{P}_0 \parallel z$-axis, the intensities of the NSF and SF channels are reversed.  In (d), the result for the (1\,\,0\,\,$1+\delta$) magnetic peak also indicates that FeAs has ($15 \pm 5$)\% more spin ploarization in the $b$-direction than the $a$-direction, a result consistent with those for the (0\,\,0\,\,$+\delta$) magnetic peak.  In (e) the flipping ratio (NSF/SF) of the (0\,\,0\,\,$+\delta$) reflection is shown to be temperature independent up to the N\'eel point.  The data presented in (a) through (c) were taken at 15 K.}
\label{FeAs_polarized}
\end{figure}

The crystal was first measured with the scattering vector $\mathbf{Q}$ in the (H0L) plane and then the (0KL) plane.  In the (H0L) configuration, the $b$-axis is along the $\zeta$ axis, and in the (0KL) configuration, the $a$-axis is along $\zeta$.  Figure \ref{FeAs_polarized}a shows the NSF and SF cross sections for the (0\,\,0\,\,2) nuclear reflection, which was used to obtain the NSF/SF flipping ratio ($\approx$ 15).

The measured intensities of the NSF and SF channels in the (0\,\,0\,\,$+\delta$) magnetic reflection (Figure  \ref{FeAs_polarized}b) completely rule out the possibility of a SDW model with spin polarization only in the $b$-direction.  Since the $a$-axis is parallel to $\eta$, no intensity should have been observed in the SF channel for a SDW.  Nevertheless, the spin amplitude is different in the $b$-direction from the $a$-direction as evidenced by the small difference between the NSF and SF intensities of the (0\,\,0\,\,$+\delta$) magnetic reflection (Figure  \ref{FeAs_polarized}b and c).  Measurements of this reflection in the (H0L) and (0KL) planes and the (1\,\,0\,\,1$+\delta$) magnetic reflection (Figure \ref{FeAs_polarized}d) consistently show that the spin amplitude is larger in the $b$-direction than the $a$-direction.  Averaging the spin-flip channels and non-spin flip channels for fourteen measurements in the (H0L) plane and two in the (0KL) plane, the spin amplitude in the $b$-direction was found to be ($15 \pm 5$)\% stronger than in the $a$-direction.  This ratio was measured up to 70 K and is temperature independent within error (Figure \ref{FeAs_polarized}e).  The inequivalence between the $a$ and $b$-components of spin polarization are consistent with the anisotropic transport and magnetic susceptibility, \cite{segawa_2009} and suggest that the envelope of the spiral traces out an ellipse rather than a circle as illustrated in Figure \ref{FeAs_magnetism}.  Since both the direction and amplitude of the spin polarization vector are modulated, we term the magnetic structure a noncollinear SDW.

The polarized results pointing to an anisotropy in the ($ab$)-plane can also be modelled equally well by a canting along the $c$ axis.  The unpolarized diffraction work carried out on BT1 and BT9 are also consistent with a small canting (on the order $\approx 10^{\circ}$), though the powder and single crystal fits produce an error bar larger than this value.  However, the susceptibility measurements displayed in Ref. \onlinecite{segawa_2009} find no kink in the susceptibility along the $c$ axis pointing to the absence of ordering along this direction and suggesting the magnetic ordering is two-dimensional.  Furthermore, the critical exponents derived from the intensity as a function of temperature also point to two-dimensional ordering rather than a three-dimensional canted ellipse.  Based on these results, we believe the two dimensional non-collinear spin-density wave picture presented, describes the available data the best.

\section{Discussion and Summary}

Analysis of the temperature dependent measurements to obtain a critical exponent have assumed that the magnetic phase transition is of second-order.  However, the results of the polarized neutron work show that the spiral picture best describes the magnetic ordering from 15 K up to $T_N$.  This would imply a failure of representational analysis since according to the Landau-Lifshitz theory on second-order phase transitions, the spin density must transform through a basis within a single irreducible representation.\cite{dimmock_1963, franzen_1990}  FeAs goes through two representations, $\Gamma_3$ and $\Gamma_4$.  While in the orthorhombic $Pnma$ symmetry $\Gamma_3$ and $ \Gamma_4$ are different, upon lowering the symmetry to orthorhombic group $Pna2_1$, the two representations become degenerate.    In the lower orthorhombic setting, mirror symmetry is lost, causing $\Gamma_3$ and $ \Gamma_4$  to become degenerate.  Indeed, the exact crystal symmetry of FeAs is somewhat ambiguous as past single crystal studies have come to different conclusions on whether $Pna2_1$ or $Pnma$ is the correct space group.\cite{lyman_1984, selte_1969} 

A lowering of the crystal symmetry of FeAs, however, may not be necessary in order for $\Gamma_3$ and $\Gamma_4$ to be degenerate and therefore mix at the magnetic ordering temperature.  It is possible that the magnetic symmetry of FeAs is independent from the crystal symmetry, a strategy used by Kallel \textit{et al.} in their calculations of the propagation vectors for FeP, CrAs, and MnP, all of which have the crystal structure of FeAs and show similar modulated magnetism.\cite{kallel_1974}  Furthermore, the two representations could be close enough in energy that they are not sufficiently separated in temperature to be distinguishable in our temperature-dependent work.

The elliptical model presented here may help to explain the the unusual transport properties presented in Ref. \onlinecite{segawa_2009}.   The anisotropy in the susceptibility is naturally explained by the elliptical model.  The distorted ellipse may also explain the difference between the field cooled and non-field cooled susceptibility along the $b$-axis.  The unusual change in the Hall number at low temperatures cannot be explained naturally with this model and future theoretical studies are required to explain this.

It is interesting to compare the magnetic structure of FeAs to other closely related systems to understand the possible exchange mechanisms leading to the non-collinear SDW.   In structurally related MnP, several magnetic ground states, both collinear and noncollinear compete to form rather rich magnetic phase diagrams.\cite{becerra_1980, shapira_1981}  Although the consensus on the ground state of MnP is that of a spiral phase, the polarized neutron work of Forsyth \textit{et al.} also found that the spiral traced out an ellipse rather than a circle.\cite{forsyth_1966}  In fact, the ratio of spin amplitude in the $a$-direction to the $b$-direction in MnP was also found to be $\approx 0.8$, strikingly similar to the ratio we observed in FeAs (See Figure \ref{FeAs_polarized}e).  This interesting find has seemingly gone unnoticed in most of the literature on MnP.  

Several theoretical studies have been undertaken to understand the exchange interactions in these transition metal pnictides leading to the observed modulated magnetism.  In the study by Takeuchi and Motizuki,\cite{takeuchi_1967} seven isotropic exchange parameters in the Heisenberg model were necessary to stabilize the spiral phase in MnP.  Likewise, Kallel \textit{et al.} required several isotropic exchange parameters to elucidate the magnetic phase diagrams of MnP, FeP, and CrAs.  In addition, the right magnetic structure and propagation vectors were also calculated by including both symmetric and antisymmetric interactions among first nearest neighbors.  Here again, lowering of the magnetic symmetry is necessary so that representations $\Gamma_3$ and $\Gamma_4$ (the present paper's notation) are degenerate.  This in turn allows the antisymmetric interaction of the Dzyaloshniskii-Moriya type to be included.   The authors, however, found this antisymmetric interaction to be abnormally large--even larger than the isotropic exchange for the case of CrAs.   Since these are all transition metal compounds where orbital contributions should be comparatively small, this result was found to be implausible by the authors.\cite{kallel_1974}

The calculations by Dobrzynski and Andresen, however, showed that the antisymmetric interaction found by Kallel \textit{et al.} for the case of MnP is unnecessary if more long-range isotropic exchange interactions are taken into account (\textit{i.e.} more nearest neighbor shells included within a Heisenberg model).\cite{dobrzynski_1989}  A more rigorous treatment of the long-range interactions in these materials was done by Sj\"ostr\"om with use of band theory to study the magnetic ground states in MnP, FeP, MnAs and CrAs.\cite{sjostrom_1992}  Sj\"ostr\"om included aniostropic exchange, both symmetric and antisymmetric, in addition to the isotropic exchange terms to find the right ground state among the possible models, which included the spiral, SDW, and ferromagnetic states.  In all but MnP, was the right ground state calculated.  Interestingly for MnP, Sj\"ostr\"om found the ferromagnetic state to have the largest lowering of the anisotropic energy (as expected for aligning along the easy axis).  MnP changes from a spiral state to a ferromagnetic state at 47 K with the moment pointing in the $b$-direction (in $Pnma$ setting).\cite{becerra_1980}

Unfortunately, no calculations have been performed on FeAs, possibly due to the lack of accurate information regarding its magnetic structure.  Moreover, none of the theoretical papers on the modulated magnetism of metal pnictides have addressed the possibility of the spiral tracing out an ellipse instead of a circle as was found for FeAs in the present work and MnP by Forsyth \textit{et al}.\cite{forsyth_1966}  This phenomenon, however, has been observed before for Gd compounds.   Rotter \textit{et al.} have termed this structure as noncollinear amplitude-modulated (NCAM) antiferromagnetism and have used it to successfully explain an anomaly in the specific heat of GdCu$_2$ near its magnetic transition.\cite{rotter_2001}  The anisotropy of the exchange interactions is implicated in the NCAM ordering, and dipole-dipole interactions found to dominate the anisotropic term for a host of Gd compounds.\cite{rotter_2003}  While these dipole-dipole interactions could play a role in FeAs, there are some key differences with rare-earth based systems that should be taken into account. In FeAs, the moment size is comparatively smaller ($\approx$ 0.5 $\mu_B$) and more delocalized as evidenced by the drop-off of the magnetic reflection intensities with higher scattering angle.  The work by Sj\"ostr\"om has shown that the effects of the conduction electrons in transition metal pnictides is not negligible, affecting the anisotropic exchange terms. 

In summary, we have demonstrated that unpolarized powder techniques are not sufficient for distinguishing the difference between the possible modulated magnetic structures of FeAs.  Since the critical exponent of FeAs is similar to the K$_2$NiF$_4$ and K$_2$MnF$_4$ systems, the magnetic interactions in FeAs would be best described by a two-dimensional Heisenberg model.  This is consistent with a spiral magnetic structure since it constrains the moment to lie in the $ab$-plane.  Polarized neutron diffraction revealed that anisotropy exists in this system with 15(5)\% more spin polarization in the $b$-direction than in the $a$-direction. As such, the magnetic structure of FeAs should be thought of as modulated in both spin amplitude and direction. Given the small moment size of $\approx 0.5 \mu_B$ and the temperature behavior of the magnetization near the N\'eel temperature, the ordering in FeAs is more accurately described as a non-collinear spin density wave.  

Modulated antiferromagnetism has been well documented in elemental metals such as Cr and the rare earths,\cite{fawcett_1968, elliott_1964} but the arsenides resist straightforward applications of the theories explaining the magnetic ground states of those elements because of the role that the mediating anions have on the exchange interactions. Thus, the possibility of itinerant, direct, and super-exchange interactions should exist in these materials.  Surely, FeAs merits more careful attention due to its similarity in bonding interactions with the FeAs-based superconductors.  In our powder results for FeAs, the Fe--Fe bond distances were found to be 2.919(1) \AA and 2.797(1) \AA, similar to those of the superconducting parent phases BaFe$_2$As$_2$ ($\approx 2.80$ \AA),\cite{rotter_2008b} LaOFeAs ($\approx 2.85$ \AA),\cite{kamihara_2008} and NaFeAs (($\approx 2.79$ \AA).\cite{parker_2009}   Thus, more work on FeAs, including theoretical and inelastic neutron spectroscopy, should further elucidate magnetic interactions in these increasingly important systems.

This work was supported by AFOSR-MURI Grant No. FA9550-09-1-0603.
%

\end{document}